# Geometrical holographic display

Guangjun Wang *

Jingmen Tanmeng Technology Co.,Ltd., Jingmen, Hubei 448000, China

*Corresponding author: wgj@tmkjg.com

**Abstract**

Is it possible to realize a holographic display with commercial available components and devices? Is it possible to manipulate light to reconstruct light field without using coherent light and complicated optical components? Is it possible to minimize the amount of date involved in building 3D scenes? Is it possible to design a holographic video display that doesn't require huge computational cost? Is it possible to realize a holographic video display with portable form like a flat-panel display commercialized nowadays? This research gives yes answers to all the above questions. A novel geometrical holographic display was proposed, which uses geometrical optical principle to reproduce realistic 3D images with all human visual cues and without visual side effects. In addition, a least necessary light field representation was introduced which can provide guidance for how to minimize the amount of data involved when designing a FP3D or building 3D scenes. Finally, a proof-of-concept prototype is set up which can provide true 3D scenes with depth range larger than 5m.

## 1. Introduction

Three-dimensional (3D) display technology is becoming popular and popular and a lot of kinds of solutions for 3D display have been proposed. The most popular one(e.g. 3D display in cinema) is the binocular stereoscopic display techniques[1-4]. Under the scheme of binocular parallax theory, two perspective views are provided to two pupils of the viewer separately. The viewer gets a 3D sense through the convergence of the light beams from the two perspective views. However, to see the corresponding perspective views clearly, the eyes have to focus on the screen. Thus, a discrepancy between the convergence distance and the focusing distance exists. This kind of accommodation-convergence conflict violates humen's physiological habits and causes visual fatigue to the viewers[5]. An ideal 3D display is one that can provide all human visual cues and does not conflict with the human 3D perception so that it is free from visual fatigue. In order to show a scene in its 3 dimensions, researchers have proposed many solutions to realize ideal(or real) 3D display[6-13]. These 3D display technology can be divided into two categories: volumetric 3D displays and flat panel form 3D display. However, none of the proposed solutions are particularly well suited to daily living display devices (television, tablets or mobile phones), which require the combination of a portable form factor, a high spatial resolution and real light field reconstructing[3]. Volumetric 3D displays are relatively mature technologies, but they are large in size and even contain dangerous high speed moving parts[14]. What's more, the 3D scene provided by volumetric 3D display usually looks subtranslucent, thus not quite realistic enough.

Flat panel form 3D display(FP3D), represented by computer-generated holography [3,5,16], can be designed into a compact and light form factor, which means it is more suitable for daily living display devices. However, there are still many fundamental limitations for realizing a FP3D. To build a FP3D, the following research challenges need to be tackled.

First of all, the hardware for FP3D is very complex and sophisticated. The necessary cues needed for 3D perception include both monocular cues(accommodation, motion parallax, occlusion and shading) and binocular cues(vergence and stereopsis)[10,17]. Only when the display device can reconstruct wave front generated by objects in free space can it account for all the visual cues. This means it must have the ability to manipulate light beams, then it can construct the desired optical light fields—ideal 3D scene. If we want to manipulate light beam, we must design a special kind of pixel that can control color and intensity and light emit direction. This is quite a difficult task because it needs an extra degree of freedom than conventional pixels used in 2D display panels(e.g. LCD), which can only control color and intensity. Computer generated hologram(CGH) displays make use of interference phenomenon and diffraction phenomenon to control the direction of light[16]. According to wave optic theory, in order to modulate the wavefront of light efficiently, the size of pixels of spatial light modulator (SLM) must close to wave length(<1um)[17]. However, it is almost impossible to design and fabricate a large active area (several square decimetre) SLM with pixels size less than 1 um. Another disadvantage of CGH is that it needs coherent source which has not yet reached the requirement of commercial application. And highly coherent light sources give rise to speckle noise. In addition, the laws of diffraction are wavelength-dependent and colors are not interchangeable, which makes it a complicated and cumbersome work to display multi-colour. Integral imaging is another most attractive and convenient method in three-dimensional display field[18,19], which uses micro-lens array to control light direction. However, it also suffers from some inherent drawbacks in terms of viewing parameters, such as viewing angle, resolution, and depth range, due to the limited resolution of the elemental image and lens array itself[3]. Many researchers work on improving the CGH based 3D display or integral imaging based 3D display[16,20-22]. There is also some artful novel design. David proposed a wide-angle novel FP3D system by introducing a multi-directional backlight[8]. He use spatial multiplex method to control the direction of light beam, hence, the spatial resolution was limited.

Secondly, it is inextricably involved in huge data volumes processing to operate a FP3D system. In a FP3D system, there is a many-to-many relationship between the scene point and light operating unit (LOU). As shown in fig.1., every LOU (or pixel) has contributions to a reconstructed scene point A in space and every scene point contains the component information of a LOU (or pixel) B. This poses huge challenges in generating, storing and transmitting of huge data volumes. Thus, it is much more calculation intensive than classic image rendering: every point in the scene can potentially affect a LOU (or pixel). This many-to-many relationship compounded with complex mathematical transformations is too costly to compute by brute-force[17].

In this research, a novel light field display is demonstrated for the first time. Similar to a hologram display, it can also reproduce realistic three dimensional images with all human visual cues and without visual side effects. The difference is that the working principle of it is based on geometrical optical principal. Therefore, there is no need to use coherent light source, which in turn brings much convenience. Thus, we name it as geometrical holographic display (GHD). The optical system is carefully designed, which use pixels as direct operating unit to operate light beams indirectly. In addition, a least necessary light field representation was introduced which can provide guidance for how to minimize the amount of data involved when designing a FP3D or building 3D scene. Finally, a proof-of-concept prototype is set up which can provide true 3D scenes with depth range larger than 5m.

## 2. Design

### 2.1 Design consideration

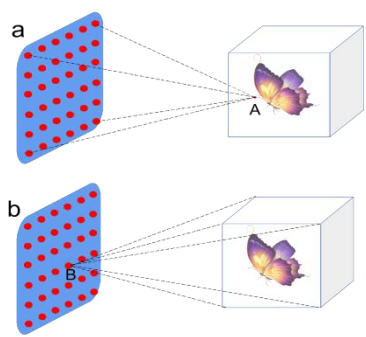

Fig. 1. The many-to-many relationship between the scene point and light operating unit (LOU) in light field display.

a) Every LOU (or pixel) has contributions to a reconstructed scene point A in space.

b) Every scene point contains the component information of a LOU (or pixel) B.

The first question that needs to be considered is whether it is possible to reduce or minimize the huge dataset, which is caused by the many-to-many relationship shown in Fig.1a and Fig.1b, of light field representation? In order to simplify the light field representation, levoy proposed a 4D light field representation[23]. As shown in figure 2a, the solution they propose is to parameterize lines by their intersections with two planes(uv plane and st plane) in arbitrary position, and don't care about lines those parallel to the planes. The great advantage of this representation is that it contains all the vision information even though it lost partial information of the complete light field (the red dashed in figure 2a). But 4D light field representation is not simple enough, it still redundancy to some extent (e.g. the red dashed not parallel to the planes in figure 2b). Evidently, the lines those not parallel to the planes but can't enter pupils can't provide any visual information either. Thus, we introduce a least necessary light field representation by removing all the lines which have no visual contribution. As shown in figure 2b, by constraining the active area of st plane to pupil zone, we get the least necessary light field representation, which is written as $L = (u,v,s,t \mid s,t \in \text{pupil})$. Even though it only takes the beams that enter pupils into consideration, it can provide all visual cues. Above all, this representation can minimize the data amount involved in 3D systems. The only drawback of this representation is that it needs tracking the viewer's eyes to realize a large viewing angle. In fact, there were some researchers who had made use of the least necessary light field representation to design novel and practical 3D systems unwittingly[16,20,22,24].

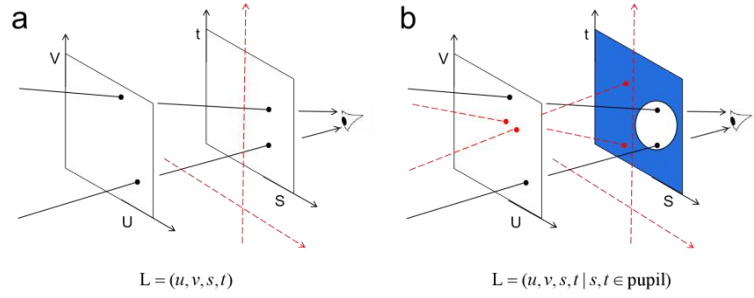

Fig. 2. Schematics of light field representation.

a) 4D light field representation: parameterize lines by their intersections with two planes(uv plane and st plane).

b) Least necessary light field representation: only take the lines those enter pupil into consideration.

In addition to the optimization of basic principles, the hardware system also needs to be optimized for real 3D display. In order to reconstruct light field we must try to control light beam, but people are only familiar with controlling pixels. It is difficult to design and manufacture LOU that can control beam directly and there is no efficient way to control beam directly. A possible solution is that by using pixel as operating unit and introducing a special kind of light transform method to control beam indirectly. Another question is that even the least necessary light field representation can reduce the amount of data for 3D scenes. It still requires huge computational cost in a real 3D display when compared with 2D displays. If we can use physical process to replace coding/decoding computation process (e.g. interferogram calculation process in CGH system), a significant reduction in computational cost can be achieved. Finally, from the perspective of user/customer, the basic requirement is that the display system must be nifty portable, which means both light in weight and small in size. In fact, folding or scrolling form screen are more popular and only FP3D has possibilities to satisfy these requirements.

## 2.2 Geometrical holographic display

In our previous works[24-26] we proposed a real time reconstructed hologram display system which uses optical process to generate interference pattern of object light and reference light on a rear projection screen directly (instead of generating interference pattern by computing process). This novel design solved almost all the problems mentioned above: it without any complex and sophisticated LOU and uses optical process to replace computation process. The only drawback is that it still needs coherent light which has not yet reached the requirement of commercial application. So we give an improved design, geometrical holographic display, based on real time reconstructed hologram display system. Fig.3a. shows the structure of GHD and its working principle. The GHD is composed of a geometrical holographic lens and a volumetric projector and it looks like a lamp with a screen. Similar to previous works[26], the projector uses micro volume scan method to generate original 3D object, then the outgoing light from the volumetric projector forms an enlarged 3D virtual object in free space. But the enlarged 3D virtual object can't be seen directly because it is divergent. The geometrical holographic lens, as shown in Fig.3b. , contains two array of strip mirrors which form a cross structure. This special structure is similar to transmissive mirror device[27]. Light emits from an object point O can form an image point at the symmetrical position after reflected by the first mirror array and the second mirror array successively. In GHD, the geometrical holographic lens is used to transform the divergent 3D virtual object to a convergent one which is visible. Compared to real time reconstructed hologram display[26], the great improvement of GHD is that it introduced a passive optical component, geometrical holographic lens, to realize image transformation rather than using diffraction and interference optical process. The special system allows us to use geometrical optical methods to reconstruct true 3D scenes(with all visual cues), so we name it geometrical holographic display. It should be noted that it needs eye tracking system to realize a large viewing angle because the visual window, which is located at the conjugate position of volumetric projector, is narrow. In fact, eye tracking requirement is a common feature of FP3D display which uses the least necessary light field representation[16,20].

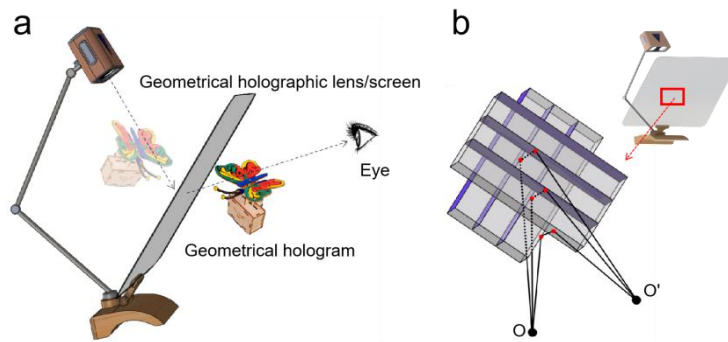

Fig. 3. The structure of GHD and its working principle.

a) The GHD is composed of a geometrical holographic lens and a volumetric projector and it looks like a lamp with a screen.

b) Structure of the geometrical holographic lens: it contains two array of strip mirrors and forms a cross structure.

## 3. Experiment

Two experiments are conducted to validate the GHD. First, the occlusion and shading effect is checked by placing virtual images in the air with different depths. In this experiment, a static volumetric projector was introduced. The static volumetric projector can project three static images in free space which are located at different depths. Second, a full-screen colour video is played. In this experiment, a commercial 2D projector was used to replace the volumetric projector. By tuning the focus ring manually, the video can be displayed at different depths too, even though it can't provide 3D information directly.

## 4. Result and discussion

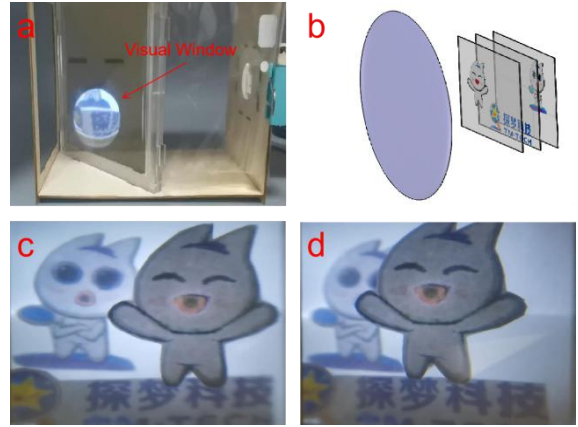

Fig. 4. Static volumetric 3D image.

a) Picture of the visual window.

b) Schematics of static volumetric projector.

c), d) Picture taken at different view angles.

The circular bright zone, shown in Fig. 4a, is the visual window. Once an eye or a camera enters this zone, the displayed scene can be seen. Fig. 5a illustrates the optical principle, a displayed point "O", which is the conjugate of "O'" projected by the static volumetric projector, is located physically at the spatial position where it is supposed to be and emits light from that position to form a real image in the eyes of viewers. This novel display mechanism allows it to display object point on the screen (geometrical holographic lens) and before the screen (O1) and behind the screen (O2), as shown in Fig. 5b. This is quite different from a 2D display, which can only display object points on the screen (O3 and O4 in Fig. 5c). That is why it can provide both physiological and psychological depth cues to the human visual system to perceive 3D objects. As shown in Fig. 4, the displayed static objects are placed along depth direction. By changing the angle of the cellphone camera, different images, which show obvious occlusion and shading phenomenon, are captured as shown in Fig. 4c, d. The movie clip is available in Supplementary Movie 1.

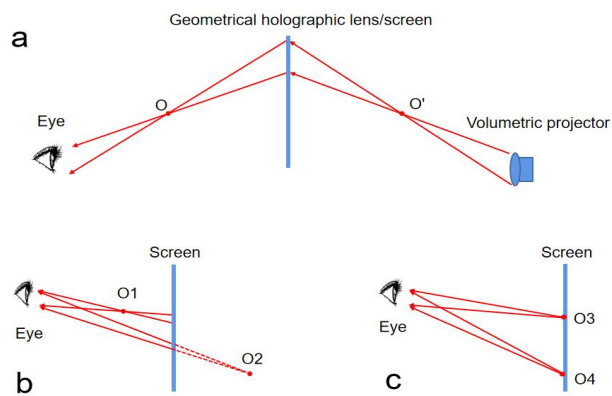

Fig. 5. Optical principle of GHD.

a) A displayed point "O", which is the conjugate of "O'" projected by the volumetric projector, is located physically at the spatial position where it is supposed to be and emits light from that position to form a real image in the eyes of viewers.

b) Light field display: can display object point before the screen (O1) and behind the screen (O2).

c) 2D display: can only display object points (O3 and O4) on the screen.

Figure 6 shows still shots taken from the full-screen colour video display in Supplementary Movie 2. The scene is about a toy butterfly opening and closing its patterned wings. At the beginning of the movie, the focal plane of the projector was set at a near distance. The blurry picture, shown in Fig. 6a, is taken with a cellphone before focusing. Then touch the focusing button, a clear picture taken, shown in Fig. 6b. When tuning the focal plane of the projector from near distance to far distance, the clear image becomes blurry again, shown in Fig. 6c. This is because the focal plane of projector is not coincident with the focal plane of cellphone camera any more. Once refocused the cellphone on the image plane, the movie becomes clear again, Fig. 6d. This movie demonstrates the unique feature of GHD providing the correct accommodation effect as real objects do. What's more, the tuning range of the distance of focal plane, in the movie, is larger than 5 m, which has not been realized by conventional 3D displays. In fact, the theoretical depth range of GHD is infinite. In Fig. 6b and Fig. 6c, the tiny gridding of geometrical holographic lens can be vaguely seen. This is because the focus distance of the camera is too close to the geometrical holographic lens. When focus at a far distance, the gridding disappears (Fig. 6a and Fig. 6d). If the gridding is tiny enough, it will be invisible even focused on the geometrical holographic lens.

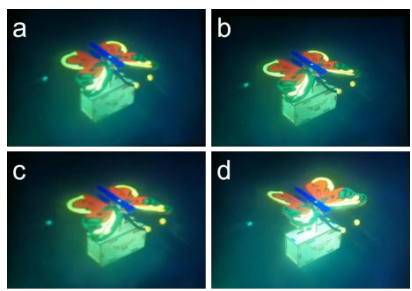

Fig. 6. Full-screen colour video: tuning the depth of image from near distance to far distance and refocusing the image.

Before tuning the depth of image a) image out of focus, b) image in focus.

After tuning the depth of image c) image out of focus, d) image in focus.

It can be seen that, even though the proof-of-concept prototype is very primitive and far from the best configuration, the displayed scene has better or comparable image quality than traditional 3D systems[6,7,9,13,16,21,28-36]. If the commercial 2D projector is replaced by a volumetric projector, as proposed in our previous works[24,26], improved display effect can be expected. The GHD system uses two ways to solve the enormous computational cost problem in CGH system. Firstly, it introduced the least necessary light field representation which can greatly reduce the data amount involved in 3D scenes. Secondly, it uses pixels as direct operating unit to operate light beams indirectly.This feature can avoid the interferogram calculation procedures in CGH systems and save computation power. What's more, the GHD system makes use of geometrical optical methods to realize the purpose of light field reconstruction, which can avoid using coherent light. This in turn avoids the difficulty of displaying color images and the speckle noise caused by highly coherent light source in CGH system.

It should be noted that even a commercial 2D projector can provide high quality image in the GHD system, it is only for validating the system and it is far from the best situation. The GHD system is so unique that there is no similar display system, thus every detail should be designed from scratch to realize a mature product. And there are many aspects that need to be balanced elaborately, e.g. light intensity, light distribution, area (and number) of visual window, configuration structure, etc, to achieve a perfect GHD. Our future work will focus on these themes. For the aspect of manufacture, GHD is much easier to manufacture than traditional CGH system, because there is no complicated and exquisite SLM and no need to use coherent light. They can be made using the existing manufacturing processes and, therefore, require little additional cost. Since the volumetric projector can be designed very small and the geometrical holographic lens can be designed into a flexible form, there is a high degree of freedom in designing the shape of a system, folding or scrolling form can be easily introduced to design and manufacture a portable/wearing device.

## 5. Conclusion

In this paper, a novel light field display named GHD is demonstrated for the first time. It uses pixels as direct operating unit to operate light beams indirectly, so there is no need to use coherent light and complicated optical components. Another benefit is that it can significantly reduce the computational cost by replacing the coding/decoding computation process with optical /physical process. All the components used in the system are easy to design and manufacture or even commercial available. The flexible optical configuration allows a fold screen form or scroll screen form which is quite suitable for portable/mobile/wearable devices. In order to minimize the amount of date involved in building 3D scenes, a least necessary light field representation was introduced. This representation can provide guidance for how to minimize the amount of data involved when designing a FP3D or building 3D scene. Finally, a proof-of-concept prototype is set up which can provide true 3D scenes with high image quality . And the depth of the scenes can be larger than 5 meters.